\documentclass[aps,amssymb,amsmath,
twocolumn,showpacs,superscriptaddress,
groupedaddress]{revtex4-1}

\usepackage{graphicx}
\usepackage{dcolumn}
\usepackage{bm} 
\usepackage[table,xcdraw]{xcolor}         
\begin{document}
\title{The use of Slater-type spinor orbitals in algebraic solution \\ of two-center Dirac equation}
\author{A. Ba{\u g}c{\i}}
\email{albagci@univ-bpclermont.fr}
\author{P. E. Hoggan}
\affiliation{Institute Pascal, UMR 6602 CNRS, University Blaise Pascal, 24 avenue des Landais BP 80026, 63177 Aubiere Cedex, France}
\begin{abstract}
The use of Slater-type spinor orbitals in algebraic solution of the Dirac equation is investigated.\\
The one- and two-center integrals constitute the matrix elements arising in generalized eigenvalue equation for one-electron atoms and molecules are evaluated over Slater-type spinor orbitals via ellipsoidal coordinates.
These integrals are calculated through numerical global-adaptive method with Gauss-Kronrod numerical integration extension.
The calculations are performed for electronic structure of ground and excited states of one-electron atoms and diatomic molecules. 
The screening constants are allowed to be variationally optimum values for given nuclear separation.
The obtained results are compaired with the results those found in the literature.
The procedures discussed in this work are capable of yielding highly accurate relativistic two-center one-electron integrals for all ranges of orbital parameters.
Besides provides an efficient way to overcome the problems that arise in relativistic calculations. 
\begin{description}
\item[Keywords]
Dirac equation, algebraic approximation, Slater-type spinor orbitals, numerical integration.
\item[PACS numbers]
... . 
\end{description}
\end{abstract}
\maketitle

\section{\label{sec:intro}Introduction}	
Molecular wave-functions are generally obtained from a linear combination of atomic orbitals (LCAO-MO) \cite{Roothaan1951} employed and studied for many years with the atomic orbitals represented by analytical basis functions.
These studies are performed in two groups of basis function: Gaussian-type functions and exponential-type functions. 
Gaussian-type functions possess a great advantage in the simplicity of evaluation of molecular integrals and are generally preferred in large scale calculations. 
Many computer programs based on Gaussian-type functions have been developed to be used in the fields of applied science. 
However, Gaussian-type functions are unable to represent the correct behavior of the wave-function at the nuclei and at large distances from it.
On the other hand, exponential-type functions are better suited then Gaussian-type functions to represent the electronic wave-function in both cases since they satisfy Kato's conditions \cite{Kato1957, Agmon1985}. \\
Two aspects of the required nonrelativistic electronic structure calculations by empliying LACO-MO method are therefore of fundamental importance. 
Firstly, the choice of basis function and secondly integral evaluation to obtain the matrix elements in the chosen basis function must be examined.

In relativistic electronic structure theory the situation is much more complicated. Although the methods have been algebraically described for many-electron systems \cite{Kim1967, Leclercq1970, Quiney1987, Koc1994} still it is contain significant problems. \\
First of all, the criteria for choosing basis functions should be reconsidered.
The numerical convergence of results to nonrelativistic limit by the use of Gaussian-type orbitals is slow or sometimes even does not lead to the right results as they do not satisfy Kato's conditions. 
Also, they are limited by quantum electrodynamics corrections \cite{Pachucki2009}. These ensure maintaining the suitability of exponential-type orbitals in theoretical investigations even they has been limited due to difficulties in efficient calculations of multicenter integrals.
On the other hand, finite-size nuclear model only requires the use of Gaussian-type orbitals \cite{Quiney1989} which should be emphasized.\\ 
The analytcal expressions by the solution of Dirac equation \cite{Grainerbook},
\begin{align}
\hat{H}_{\mathit{D}}\Psi=E\Psi
\end{align}
here,
\begin{align}
\hat{H}_{\mathit{D}}=c(\vec{\alpha}.\hat{\vec{p}})+mc^{2}\beta+V(r)
\end{align}
is one-electron Dirac operator, 
\begin{align}
\Psi=\begin{pmatrix} \psi^{L}\\\psi^{S} \end{pmatrix}
\end{align}
is two-component form of electron wave-function and
\begin{align}
\vec{\alpha}=\begin{pmatrix} 0& \vec{\sigma} \\ \vec{\sigma}& 0 \end{pmatrix}
\qquad\text{}\qquad
\beta=\begin{pmatrix} I& 0 \\ 0& {-I} \end{pmatrix}
\end{align}
with, c is speed of light, $\vec{\sigma}$ are Pauli spin matrices, $\hat{\vec{p}}$ is momentum operator, $m$ is the rest mass of electron, I is $2X2$ unit matrix, $V(r)$ is interaction potential and $\psi^{L}$, $\psi^{S}$ are represent large- and small-components of electron spinor wave-function respectively, are available only for limited interactions with strong constraint on potential \citep{Ho2006, Halberg2006}.
An approximate solution of Dirac equation gives finite-discrete energy spectrum known as positive- and negative-energy spectrum which are represent electronic, positronic energy states.
The accurate calculation of whole spectrum is quite sensitive to mathematical completeness. It is require the choice of basis functions depending on kinetic balance condition \citep{Lee1982, Stanton1984};
\begin{align}
\lim_{c \to \infty} \psi^{S}=\frac{1}{2mc}(\vec{\alpha}.\hat{\vec{p}})\psi^{L}
\end{align}
Unfortunately, expectation from obtained positive-energy eigenvalues by the solution of Dirac equation in the nonrelativistic limit to converge to the eigenvalues obtained by the solution of Schr\"{o}dinger equation may not comes true when finite-basis approximation is used.
The kinetic balance condition guarantees the implementation of the variation principle only if the basis-set used in Dirac equation matches with the basis-set used in Schr\"{o}dinger equation in the nonrelativistic limit.
The variational instability, so called variational collapse \cite{Schwarz1982-1} or finite basis-set disease \cite{Schwarz1982-2}, may arise throughout algebraic solution of the one-electron Dirac equation to find the lowest eigenvalues of positive-energy states due to the Dirac-Hamiltonian is unbounded from below.
The collapse of the obtained positive-energy eigenvalues below the lowest positive-energy ground state have been serious block in relativistic electronic structure calculations and it have been studied quite intensively by many authors since it is comprehended.
Numerous approaches were proposed in these studies. 
They can be divided into two main groups; those that based to obtain Dirac-like equation where direct variational (Rayleigh$-$Ritz) procedure can be applicable and those that based on definition new variational procedure provides rigorous upper and lower bounds to positive- and negative-energy spectrum, respectively. 
The studies in first group were possible by a few ways; applying the variation method to the Dirac equation based on the modified Dirac operator \cite{Wallmeier1981, Wallmeier1984, Volkov1989, Hill1994}, transformation of the Dirac-Hamiltonian to a block diagonal form through a Foldy$-$Wouthuysen transformation method \cite{Wallmeier1983, Kutzelnigg1984, Kutzelnigg1987, Kutzelnigg1989}, applicaiton of partitioning technique to the Dirac equation \cite{Wood1985, Rutkowski1986, Hegarty1986}, elimination of the small-components in order to get effective Hamiltonian only for large-components \cite{Hegarty1987, Lenthe1995, Dyall1997, Filatov2002, Filatov2005, Zou2011} and using appropriate finite basis-set \cite{Drake1981, Dyall1984-1, Dyall1984-2, Aerts1985, Grant1986, Quiney1987, Goldman1987, Hegarty1987, Laaksonen1988, Quiney1989} where the positive-energy eigenvalues are obtained bounded from below.
The studies in second group were posssible by using particular variational procedures, Lehmann-Maehley and Minimax methods, \cite{Lehmann1949, Maehly1952, Talman1986} to solution of one-electron Dirac equation. Notice that detailed investigations of these procedures were made in \cite{Kolakowska1996, Falsaperla1997}. \\
The use of central Coulomb potential in algebraic solution of one-electron Dirac equation brings another difficulty which is show up when relativistic angular momentum quantum numbers $(\kappa)$ are positive. Here, the obtained positive-energy eigenvalue may lie in forbidden energy gap where, between lowest true positive-energy ground state and the negative-energy spectrum threshold $-2mc^{2}$. Solution of this problem analyzed through defining basis-set approximation in \cite{Drake1981, Goldman1985, Goldman1986, Goldman1988, Goldman1989} and its origin investigated in \cite{Pestka2003, Pestka2004}.
Note that, the discussions on relativistic electronic structure calculations comprehensively can be found in \cite{Grant2010, Dyall2012, Kutzelnigg2012} published, recently. 

The aim of this paper is to show that using Slater-type spinor orbitals, which are obtained here analogously to L-spinors \cite{Quiney1989, Grant2000} the Dirac equation can be solved via Rayleigh$-$Ritz method for extended basis-sets approximation without any modification on Dirac operator. The relations for relativistic one-electron molecular integrals can be obtained with compact form and calculated accurately for noninteger values of quantum numbers via molecular auxiliary functions \cite{Bagci2014-1} in ellipsoidal coordinates and suggested numerical integration method in \cite{Bagci2014-2}. \\
The calculations are performed for electronic structure of ground and excited state of one-electron atoms and diatomic molecules with single-zeta basis-set approximation for each sign $\kappa$ which is determine the symmetry of orbitals to be included in LCAO. The screening constants are allowed to be variationally optimum values. \\
This paper is proving that the nature of Dirac equation is compatible with Rayleigh$-$Ritz method for extended basis sets approximation if Slater-type spinor orbitals is used and matrix representation of Dirac equation can be solved via procedure given for solution generalized eigenvalue equation without encountering any troubles.

\section{\label{sec:intro}Slater-type spinor orbitals}
The Slater-type spinor orbitals (STSOs) which can be consider as relativistic analogues of Slater-type functions (STFs) have the functional form of the most nearly nodeless L-spinors characterized by minimum value of radial quantum numbers $n_{r}$ \cite{Grant2007}, where $\left(n_{r}=n-\vert \kappa\vert \right)$ \cite{Landau1977} and $n$ principal quantum number. \\ 
The STSOs used in this paper determined as,
\begin{align} \label{eq:STSOs}
\chi_{\gamma\kappa\mu}^{\beta}\left(\zeta, \vec{r}\right)
=\left\{{A_{\kappa}^{\beta}r^{\gamma}+\zeta B_{\kappa}^{\beta}r^{\gamma+1}}\right\}e^{-\zeta r} 
\Omega_{\kappa\mu}^{\beta} \left(\theta, \vartheta\right)
\end{align}
where, $\kappa=\mp1, \mp2, \mp3, ...$, $\vert\kappa\vert+\frac{1}{2}\leq\mu\leq\vert\kappa\vert-\frac{1}{2}$, $\beta=\mp1$ are represent large- and small-components of STSOs, respectively. The $\Omega_{\kappa\mu}^{\beta}$ are the spin $\frac{1}{2}$ spinor spherical harmonics,
\begin{align} \label{eq:SSHs}
\Omega_{\kappa\mu}^{\beta} \left(\theta, \vartheta\right)
=\sum_{\sigma=\frac{\beta}{2}}^{} C_{\mu-\sigma \sigma \mu}^{l \frac{1}{2} j} Y_{l\mu-\sigma}\left(\theta, \vartheta\right)\phi_{\sigma},
\end{align}
\begin{align} \label{eq:SpinM}
\phi_{\frac{1}{2}}=
\begin{pmatrix}
  0 \\ 1
 \end{pmatrix},
 \quad\quad
 \phi_{-\frac{1}{2}}=
\begin{pmatrix}
  0 \\ 1
 \end{pmatrix}
\end{align}
here, the quantities C are Clebsch-Gordon coefficients. Using the explicit form \cite{Landau1977, Szmytkowski2007} of the Clebsch-Gordon coefficients the spinor spherical harmonics can be obtained by following formula,
\begin{multline} \label{eq:OSSHs}
\Omega_{\kappa\mu}^{\beta} \left(\theta, \vartheta\right)
=\frac{1}{\sqrt{2\beta\kappa+1}} \\
\times \begin{bmatrix}
sgn\left(-\beta\kappa \right) \sqrt{\left(\beta\kappa\right)+1/2-\mu}Y_{l_{\beta}\mu-\frac{1}{2}}\left(\theta, \vartheta\right) \\ \sqrt{\left(\beta\kappa\right)+1/2+\mu}Y_{l_{\beta}\mu+\frac{1}{2}}\left(\theta, \vartheta\right)
\end{bmatrix}
\end{multline}
$Y_{lm}$ are the complex spherical harmonics \cite{Condon1970} with,
\[ l_{\beta} = \left\{ \begin{array}{ll}
\beta\kappa & \mbox{\quad $\beta\kappa > 0$} \\
-\beta\kappa-1 & \mbox{\quad $\beta\kappa < 0$}, \end{array} 
\right. \]
$Y^{*}_{lm}=Y_{l-m}$ \citep{Steinborn1973}. Notice that the spinor spherical harmonics are satisfy the orthogonality relations,
\begin{align} \label{eq:ANGORT}
\int_{0}^{\pi} \int_{0}^{2\pi}
\Omega_{\kappa\mu}^{\beta} \left(\theta, \vartheta\right)
\Omega_{\kappa'\mu'}^{\beta'} \left(\theta, \vartheta\right)
d\Omega=\delta_{\beta\beta'}\delta_{\kappa\kappa'}\delta_{\mu\mu'},
\end{align}
$d\Omega=sin\theta d\theta d\vartheta$ and the operator $\hat{K}_{\theta\vartheta}$, 
\begin{align} \label{eq:KOPT}
\hat{K}_{\theta\vartheta}
\equiv \left(\hat{\sigma}.\hat{r} \right)
=\begin{bmatrix}
Cos\theta &Sin\theta e^{-i\vartheta} \\
Sin\theta e^{i\vartheta} &-Cos\theta
\end{bmatrix}
\end{align}
changes their parity since it is odd of parity,
\begin{align} \label{eq:PARITY}
\hat{K}_{\theta\vartheta}\Omega_{\kappa\mu}^{\beta} \left(\theta, \vartheta\right)
=-\Omega_{\kappa\mu}^{-\beta} \left(\theta, \vartheta\right).
\end{align}
The coefficients $A_{\kappa}^{\beta}$, $B_{\kappa}^{\beta}$ included in the radial part of STSOs are defined as follows,
\begin{align} \label{eq:AC}
A_{\kappa}^{\beta}
=\left(\frac{\beta \kappa}{2\gamma}\right)-\frac{1}{2} \left(1+\beta\kappa+\frac{\beta\kappa}{\gamma}-\beta N_{\kappa}^{\gamma}\right)\delta_{\vert \kappa\vert \kappa},
\end{align}
\begin{align} \label{eq:BC}
B_{\kappa}^{\beta}
=-\beta \left(\frac{N_{\kappa}^{\gamma}-\kappa}{2\gamma+1}\right),
\end{align}
where,
\begin{align} \label{eq:NK}
N_{\kappa}^{\gamma}=\sqrt{\kappa^{2}+\left(2\gamma+1\right)\delta_{\vert \kappa\vert \kappa}}
\end{align}
and, $\gamma \in \mathbb{R}^{+}$. It should be emphasized that the STSOs have the same form to as S-spinors \cite{Grant2007} if 
\begin{align}\label{eq:GAMMA}
\gamma=\sqrt{\kappa^2-\frac{Z^2}{c^2}}
\end{align}
with nuclear charge Z, except they are not independent for a large and small components and they satisfy the following system of differential equation:
\begin{multline} \label{eq:KB}
\frac{\partial}{\partial r}\chi_{\gamma\kappa}^{\beta}\left(\zeta, r\right)
=-\beta\frac{\kappa}{r}\chi_{\gamma\kappa}^{\beta}\left(\zeta, r\right) \\
+\left(\frac{\beta N_{\kappa}^{\gamma}-\gamma-\delta_{\vert \kappa\vert \kappa}}{r}+ \zeta \right)\chi_{\gamma\kappa}^{-\beta}\left(\zeta, r\right) 
\end{multline}

\section{\label{sec:intro}Definition and basic formulas}
The following linear combinations of molecular orbitals in terms of STSOs are used through the calculation of electronic energies of one-electron homo-nuclear and hetero-nuclear diatomic molecules:
\begin{align} \label{eq:LCAO-HOMO}
u_{i}^{\beta}=
\sum_{q}^{}\left[\chi_{q}^{\beta} \left(\zeta, \vec{r}_{a} \right)+ I\chi_{q}^{\beta} \left(\zeta, \vec{r}_{b} \right) \right]C_{qi}^{\beta}
\end{align}
\begin{align} \label{eq:LCAO-HETERO}
u_{i}^{\beta}=
\sum_{q}^{}\left[\chi_{q}^{\beta} \left(\zeta_{a}, \vec{r}_{a} \right)C_{qi}^{\beta a}+ I\chi_{q}^{\beta} \left(\zeta_{b}, \vec{r}_{b} \right)C_{qi}^{\beta b} \right]
\end{align}
where $q=\kappa\left(-N_{q}\leq\kappa\leq N_{q}\right)$, $i=\vert\kappa\vert\left(1\leq\vert\kappa\vert\leq 2N_{q}\right)$, $I=\mp1$ denote the gerade, ungerade states, respectively and $N_{q}$ is the upper limit of summation. The orbital parameters are chosen depending on $\kappa$ as follows,
\[ \zeta_{a}=\zeta_{b}=\left\{ \begin{array}{ll}
\zeta & \mbox{\quad $\kappa < 0$} \\
\zeta' & \mbox{\quad $\kappa > 0$}. \end{array} 
\right. \]
The calculations are performed to be obtained the energies and linear combination coefficients by solution of following generalized eigenvalue equation in a matrix form \cite{Roothaan1951, Kim1967, Leclercq1970, Koc1994, Quiney1987}
\begin{align} \label{eq:GEVE}
H'_{D}C=SCE
\end{align}
here,
\begin{align} \label{eq:OVERLAPM}
S= \begin{pmatrix}
S_{pq}^{\beta\beta} &0 \\ \\
0 & S_{pq}^{-\beta-\beta}
\end{pmatrix}
\quad\quad
C= \begin{pmatrix}
C_{pq}^{\beta} \\ \\ C_{pq}^{-\beta}
\end{pmatrix},
\end{align}
\begin{align} \label{eq:HAMILTONIANM}
H'_{D}= \begin{pmatrix}
V_{pq}^{\beta\beta} &cT_{pq}^{\beta-\beta} \\ \\
cT_{pq}^{-\beta\beta} & -2mc^2S_{pq}^{-\beta-\beta}+V_{pq}^{-\beta-\beta}
\end{pmatrix} .
\end{align}
The one-electron integrals arise in Eq. (\ref{eq:GEVE}) are defined as:\\
overlap integrals,
\begin{multline} \label{eq:OVERLAP}
S_{\gamma\kappa\mu,\gamma'\kappa'\mu'}^{\beta\beta'} \left(\vec{p},t \right) \\
=\int_{}^{}\chi_{\gamma\kappa\mu}^{\beta}\left(\zeta,\vec{r}_{a} \right)\chi_{\gamma'\kappa'\mu'}^{\beta'}\left(\zeta',\vec{r}_{b} \right)dV,
\end{multline}
nuclear attraction integrals,
\begin{multline} \label{eq:NUCLEARABA}
^{abb}V_{\gamma\kappa\mu,\gamma'\kappa'\mu'}^{\beta\beta'} \left(\vec{p},t \right) \\
=\int_{}^{}\chi_{\gamma\kappa\mu}^{\beta}\left(\zeta,\vec{r}_{a} \right)\frac{1}{{r}_{b}} \chi_{\gamma'\kappa'\mu'}^{\beta'}\left(\zeta',\vec{r}_{b} \right)dV,
\end{multline}
\begin{multline} \label{eq:NUCLEARAAB}
^{aab}V_{\gamma\kappa\mu,\gamma'\kappa'\mu'}^{\beta\beta'} \left(\vec{p},t \right) \\
=\int_{}^{}\chi_{\gamma\kappa\mu}^{\beta}\left(\zeta,\vec{r}_{a} \right)\frac{1}{{r}_{b}} \chi_{\gamma'\kappa'\mu'}^{\beta'}\left(\zeta',\vec{r}_{a} \right)dV,
\end{multline}
and, kinetic energy integrals,
\begin{multline} \label{eq:KINETIC}
T_{\gamma\kappa\mu,\gamma'\kappa'\mu'}^{\beta\beta'} \left(\vec{p},t \right)\\
=\int_{}^{}\chi_{\gamma\kappa\mu}^{\beta}\left(\zeta,\vec{r}_{a} \right)(\hat{\sigma}.\hat{\vec{p}}) \chi_{\gamma'\kappa'\mu'}^{\beta'}\left(\zeta',\vec{r}_{b} \right)dV,
\end{multline}
where, $\vec{p}=\frac{\vec{R}}{2}(\zeta+\zeta')$, $t=\frac{t^{-}}{t^{+}}=\frac{\zeta-\zeta'}{\zeta+\zeta'}$ and $\vec{R}=\vec{r}_{a}-\vec{r}_{b}$ is the inter-nuclear distance vector.
The vectors $\vec{r}_{a}$, $\vec{r}_{b}$ are radius vectors of electrons with respect to nuclear labels $a, b$.

\section{\label{sec:intro}Evaluation of one-electron molecular integrals}
In order to derive the two-center one-electron integrals it is utilizing from expansion formula for spinor spherical harmonics with same and different centers and molecular auxiliary functions \cite{Bagci2014-1, Bagci2014-2} in lined-up coordinate systems via an ellipsoidal coordinates ($\xi, \nu, \vartheta$), respectively
\begin{align} \label{eq:EXPASSHSC}
\Pi_{\kappa\mu,\kappa'\mu'}^{\beta\beta',k} \left (\theta, \vartheta \right)
=\sum_{kLM}{C_{\kappa\mu}^{\beta k}C_{\kappa'\mu'}^{\beta'k}C_{\kappa\mu,\kappa'\mu'}^{\beta\beta',k;L}Y_{LM}^{*}}\left (\theta, \vartheta \right),
\end{align}
\begin{multline} \label{eq:EXPASSHDC}
\Pi_{\kappa\Lambda,\kappa'\Lambda'}^{\beta\beta',k} \left (\xi, \nu, \vartheta \right) \\
=\frac{1}{2\pi}{C_{\kappa\Lambda}^{\beta k}C_{\kappa'\Lambda'}^{\beta'k}C_{\kappa\lambda,\kappa'\lambda'}^{\beta\beta',k}}\left (\xi, \nu \right)e^{i(\lambda-\lambda')\vartheta},
\end{multline}
\begin{multline} \label{eq:AUXILIARYG}
^{P_{1},Q_{1}}\mathcal{G}_{\sl N_{2} \sl N_{3} \sl N_{4}}^{{\sl N_{1}},q} \left(p_{1},p_{2},p_{3} \right) \\
=\frac{p_{1}^{\sl N_{1}}}{\left({\sl N_{4}}-{\sl N_{1}} \right)_{\sl N_{1}}}
\int_{0}^{1}\int_{-1}^{1}{\left(\xi\nu \right)^{q}\left(\xi+\nu \right)^{\sl N_{2}}\left(\xi-\nu \right)^{\sl N_{3}}}\\
\times \begin{bmatrix}
P_{1}\left[{\sl N_{4}-N_{1}},p_{1}\left(\xi+\nu \right) \right] \\ Q_{1}\left[{\sl N_{4}-N_{1}},p_{1}\left(\xi+\nu \right) \right]
\end{bmatrix}e^{p_{2}\xi-p_{3}\nu}d\xi d\nu
\end{multline}
here, $\left(a \right)_{n}$ is the Pochhammer symbol, $P\left[\alpha,x \right]$ is the normalized incomplete gamma, $Q\left[\alpha,x \right]$ is normalized complementary incomplete gamma functions, $C_{\kappa\mu}^{\beta k}$ are the Clebsch-Gordan coefficients, $C_{\kappa\mu,\kappa'\mu'}^{\beta\beta',k;L}$ are the Gaunt coefficients and they defined by following form;
\begin{multline} \label{eq:INTCGC}
C_{\kappa\mu}^{\beta k}
=(-1)^{\dfrac{2\delta_{k\vert\beta\kappa\vert,k\beta\kappa}+ \left[\vert{\mu-\frac{k}{2}\vert}-\left(\mu-\frac{k}{2} \right) \right]}{2}} \\
\times\sqrt{\frac{\beta\kappa+\frac{1}{2}-k\mu}{2\beta\kappa+1}}
\end{multline}
\begin{multline} \label{eq:GAUNTC}
C_{\kappa\mu,\kappa'\mu'}^{\beta\beta',k;L} \\
=\sqrt{\frac{2L+1}{4\pi}}C^{L}\left(l_{\beta}\mu-\frac{k}{2};l'_{\beta}\mu'-\frac{k}{2} \right)\delta_{M,\mu-\mu'}.
\end{multline}
The coefficients $C_{\kappa\lambda,\kappa'\lambda'}^{\beta\beta',k}\left (\xi, \nu \right)$ are the product of two normalized associated Legendre functions in ellipsoidal coordinates and they are determined as,
\begin{multline} \label{eq:LegendreP}
C_{\kappa\lambda,\kappa'\lambda'}^{\beta\beta',k}\left (\xi, \nu \right) \\
=\sum_{abc}g_{ab}^{c}\left(l_{\beta\lambda},l'_{\beta'\lambda'} \right)\times\frac{\left(\xi\nu \right)^{c}}{\left(\xi+\nu \right)^{a}\left(\xi-\nu \right)^{b}}.
\end{multline}
Here, $\Lambda=\vert\mu\vert $, $\Lambda'=\vert\mu'\vert $, $\lambda=\vert\mu-\frac{k}{2}\vert $, $\lambda'=\vert\mu'-\frac{k}{2}\vert $ and $-1 \leq k(2) \leq 1 $. Please see \cite{Guseinov1970, Guseinov1995} for the definitions Gaunt coefficients and $g_{ab}^{c}$.

Considering the relations given in Eqs. (\ref{eq:EXPASSHSC}, \ref{eq:LegendreP}) the relativistic two-center integrals over normalized STSOs are defined by following formula, \\
the overlap integrals,
\begin{multline} \label{eq:OVERLAPANALEX}
S_{\gamma\kappa\Lambda,\gamma'\kappa'\Lambda}^{\beta\beta'}\left(p,t \right)
=C_{\kappa\Lambda}^{\beta k}C_{\kappa'\Lambda'}^{\beta' k}N_{\gamma\kappa,\gamma'\kappa'}^{\beta i,\beta' i'} \left(p,t \right) \\
X_{\kappa}^{\beta i}X_{\kappa'}^{\beta' i'}S_{\gamma+i,l_{\beta}\lambda,\gamma'+i'l'_{\beta'}\lambda}\left(p,t\right),
\end{multline}
the nuclear attraction integrals,
\begin{multline} \label{eq:NUCLEARABBANALEX}
^{abb}V_{\gamma\kappa\Lambda,\gamma'\kappa'\Lambda}^{\beta\beta'}\left(p,t \right)
=\left(\frac{t^{+}}{p} \right)C_{\kappa\Lambda}^{\beta k}C_{\kappa'\Lambda'}^{\beta' k}N_{\gamma\kappa,\gamma'\kappa'}^{\beta i,\beta' i'} \left(p,t \right) \\
X_{\kappa}^{\beta i}X_{\kappa'}^{\beta' i'}S_{\gamma+i,l_{\beta}\lambda,\gamma'+i'-1 l'_{\beta'}\lambda}\left(p,t \right),
\end{multline}
\begin{multline} \label{eq:NUCLEARAABANALEX}
^{aab}V_{\gamma\kappa\mu,\gamma'\kappa'\mu'}^{\beta\beta'}\left(\vec{p},t \right)
=\sum_{kLM}^{} \sqrt{\frac{4\pi}{2L+1}}C_{\kappa\mu}^{\beta k}C_{\kappa'\mu'}^{\beta'k}C_{\kappa\mu,\kappa'\mu'}^{\beta\beta',k;L} \\
\times R_{\gamma\kappa,\gamma'\kappa'}^{\beta\beta';L} \left(p,t \right) Y_{LM}^{*} \left (\theta, \vartheta \right),
\end{multline}
and the kinetic energy integrals
\begin{multline} \label{eq:KINETICENERGYEX}
T_{\gamma\kappa\Lambda,\gamma'\kappa'\Lambda}^{\beta\beta'} \left(p,t \right) \\
=-\left\{N_{\kappa'}^{\gamma'}-\beta'\left(\gamma'+\delta_{\vert\kappa'\vert\kappa'} \right) \right \}^{abb}V_{\gamma\kappa\Lambda,\gamma'\kappa'\Lambda}^{\beta-\beta'}\left(p,t \right)\\
-t^{+}\left(1-t \right)S_{\gamma\kappa\Lambda,\gamma'\kappa'\Lambda}^{\beta-\beta'}\left(p,t \right)
\end{multline}
here, 
\begin{align} \label{eq:XBI}
X_{\kappa}^{\beta i}
=A_{\kappa}^{\beta}\delta_{i0}+B_{\kappa}^{\beta}\delta_{i1},
\end{align}
\begin{multline} \label{eq:BIGNORM}
N_{\gamma\kappa,\gamma'\kappa'}^{\beta i,\beta' i'} \left(p,t \right) \\
=\frac{1}{2^{i+i'}}\frac{\left[p\left(1+t\right)\right]^{\gamma+i+\frac{1}{2}}\left[p\left(1-t\right)\right]^{\gamma'+i'+\frac{1}{2}}}{\sqrt{Y_{\gamma\kappa}^{\beta}Y_{\gamma'\kappa'}^{\beta'}}},
\end{multline}
\begin{multline} \label{eq:YBI}
Y_{\gamma\kappa}^{\beta}
=\left(A_{\kappa}^{\beta}\right)^2\Gamma \left[2\gamma+1 \right]+A_{\kappa}^{\beta}B_{\kappa}^{\beta}\left[2\gamma+2 \right] \\ +\left(B_{\kappa}^{\beta}\right)^2\frac{\Gamma \left[2\gamma+3 \right]}{4},
\end{multline}
and, 
\begin{multline} \label{eq:NONOVERLAP}
S_{nl\lambda,n'l'\lambda}\left(p,t \right) \\
=\sum_{abc}g_{ab}^{c} \left(l\lambda,l'\lambda \right)^{P_{1},Q_{1}}\mathcal{G}_{\sl N_{2} \sl N_{3} 0}^{0,q} \left(p,p,pt \right),
\end{multline}
with, $0\leq\left(i,i'\right)\leq1$. Finally, taking into consideration,
\begin{multline} \label{eq:r12EXP}
\frac{1}{r_{21}}=
\sum_{LM}^{}\left(\frac{4\pi}{2L+1} \right) \left(\frac{r_{<}^{L}}{r_{>}^{L+1}} \right)Y_{LM}\left(\theta_{1},\vartheta_{1} \right)Y_{LM}^{*}\left(\theta_{2},\vartheta_{2} \right)
\end{multline}
the one-center potential $R_{\gamma\kappa,\gamma'\kappa'}^{\beta\beta';L}$ in Eq.(\ref{eq:NUCLEARAABANALEX}) can be determined by,
\begin{widetext}
\begin{multline} \label{eq:ONECPOT}
R_{\gamma\kappa,\gamma'\kappa'}^{\beta\beta';L} \left(p,t \right)
=N_{\gamma\kappa,\gamma'\kappa'}^{\beta i,\beta' i'} \left(1,t \right)
X_{\kappa}^{\beta i}X_{\kappa'}^{\beta' i'}\left(2t^{+} \right)\Gamma\left [\gamma+i+\gamma'+i'+L+1 \right]\frac{1}{\left(2p \right)^{L+1}} \\ 
\times\left\{P \left [\gamma+i+\gamma'+i'+L+1,2p \right] +\frac{\left(2p \right)^{2L+1}}{\left(\gamma+i+\gamma'+i'-L\right)_{2L+1}}Q \left [\gamma+i+\gamma'+i'-L,2p \right] \right\}
\end{multline}
\end{widetext}
$\vert{l_{\beta}-l'_{\beta}\vert}\leq L\left(2 \right) \leq \vert{l_{\beta}+l'_{\beta}\vert}$, $-L\leq M\leq L$.

\section{\label{sec:intro}Results and Discussions}

The studies to solve the problems arise in algebraic solution of Dirac equation were commonly carried out for hydrogenic atoms and one-electron diatomic molecules. 
Investigation of one-electron diatomic molecules or generally speaking, two-center Dirac problem over exponential-type spinor orbitals algebraically could not be performed with genuine conviction due to the difficulties in accurate calculation of two-center integrals for this group of spinor orbitals. 
The problems arise in the solution of Dirac equation also have been serious block in application of variation principle. 
Note that, some calculations were performed with Gaussian-type spinor orbitals in variational method \cite{Mark1980, Quiney2004} and Minimax method in two-center Dirac problem were investigated \cite{LaJohn1992}. But, the use of exponential-type orbitals in two-center Dirac problem via direct variational approach has been open question.
The two-center Dirac problem widely have been studied via numerical methods \cite{Laaksonen1984, Yang1991, Yu1994, Alexander1999, Kullie2001, Ishikawa2008, Artemyev2010} or perturbative treatment of relativistic effects \cite{Mark1987, Howells1990, Franke1992, Parpia1995, Tsogbayar2006} as it have not been practically possible to perform the algebraic approximations.
\\
Recently, the accuracy problem in the evaluation of molecular integrals have been solved through numerical approximation in \cite{Bagci2014-2,Bagci2014-1}. 
These improvements led to reconsider application of kinetically balanced exponential-type spinor orbitals, which obtain analogously from L-type spinor for the solution of Dirac equation in algebraic approximation. 
At first sign, the S-type spinor orbitals is known available to be used in this problem.
But, the difficulty of finding simple relations for two-center relativistic integrals still remain if large- and small-component of used spinor orbitals is not directly dependent. 
The STSOs and given relation in Eq. (\ref{eq:KB}) for their large- and small-components provide an efficient and simple way to obtain the relativistic integrals. 
Besides the S-type spinor orbitals are special case of STSOs for $\gamma=\sqrt{\kappa^2-{Z^2}/{c^2}}$.

In this study, the Eq.(\ref{eq:GEVE}) with its included matrix elements are solved for the determination of linear combination coefficients and electronic energies using Mathematica programming language. 
Schur decomposition \cite{Wolfram2003} is utilized ot obtain eigenvalues since matrix form of Dirac-Hamiltonian is not hermitian. 
The calculations are performed for ground and exicted of one-electron atoms and homo-, hetero-nuclear diatomic molecules with single-zeta basis sets approximation in linear combinations of STSOs given in Eqs. (\ref{eq:LCAO-HOMO}, \ref{eq:LCAO-HETERO}) for each sign of $\kappa$. 
Determination of nonlinear parameters have critical importance for correct representation of atomic orbitals in relativistic calculations. The screening constants are allowed to be variationally optimum values. The Powell optimization procedure \cite{Mathewsbook} is performed for defined basis sets approximation. The quantum numbers $\gamma$ are chosen to take positive integer values. Notice that, the calculations can also be performed with $\gamma=\sqrt{\kappa^2-{Z^2}/{c^2}}$ or $\gamma$ can be assigned as parameter to be optimized. Unfortunately, the robust numerical procedure given in \cite{Bagci2014-2,Bagci2014-1} for highly accurate calculation of molecular integrals is not efficient according to computational time. The analytical method which gives results for these integrals accurate as much as results given in \cite{Bagci2014-2,Bagci2014-1} should be examined eventually.
\\
The calculations for one-electron atoms are performed with integer values of $\gamma$ for arbitrary extended basis sets approximation, where $\kappa$ can take positive and negative values, without encounter any kind of problem for ground and excited states. The investigation of one-electron atoms shows that there are basis functions with integer values of quantum numbers satisfy kinetic-balance condition and they can be applicable for calculation of electronic structure of one-electron atoms. However, critical importance of two-center problem requires a detailed examination and instead of present the results obtained for one-electron atoms it is preferred to focusing on two-center Dirac problem. It is believed that, the reader have desire to see the results for one-electron atoms can perform his/her calculations easily.
\\
The results of calculations are presented in tables I-IV and figure 1, respectively. Variational stability are tested for one-electron molecules by large number of calculations. 
In these tables the results obtained for $1s\sigma_{1/2}$, $1p\pi_{3/2}$, $2d\delta_{5/2}$ electronic energy states of one-electron molecules with different values of nuclear charges and inter-nuclear distances are presented. 
The results obtained by solution of Dirac equation, where $c=137.0359895$ are given in the first row of each values of nuclear charges. In the second row the given results obtained by solution of Dirac equation in nonrelativistic limit($c=10^{6}$) or by the solution Schr\"{o}dinger equation. 
Notice that, each value given in these tables is obtained with variationally optimum values of screening constants independently.    
\\
At low values of the nuclear charges during the optimization the spurious roots are encountered for a couple values of electronic energies. The space belong to these values in tables left empty. In figure these energies are plotted depending on screening constants, where the resolution is $1/10$.
\\
In table \ref{tab:table1} the results of calculations obtained by employing minimal basis-sets approximation are presented. 
The ground and excited state electronic energies are examined in this basis sets approximation. 
The values given in second row of each nuclear charge are obtained via solution of Schr\"{o}dinger equation. 
It can be seen from this table the suggested basis functions are available to perform the calculations with minimal basis sets for any energy state with arbitrary nuclear charge without hesitate about interval initially given for optimization of screening constant.
\\
In tables \ref{tab:table2}, \ref{tab:table3} the results of calculations obtained by employing extended basis-sets approximation are presented, where upper limit of summations are chosen to as $1$ and $2$ for tables \ref{tab:table2}, \ref{tab:table3}, respectively. 
The upper limit of summation to be $1$ define two basis functions in each atom with $\kappa=-1$ and $\kappa=1$ and if it is $2$ define four basis functions in each atom with  $\kappa=-1,1$ and $\kappa=-2,2$.
In table \ref{tab:table2} the second row for given each nuclear charges are obtained by solution of Dirac equation in nonrelativistic limit and in table \ref{tab:table3} are obtained by solution of Schr\"{o}dinger equation using basis sets approximation given in \cite{Bagci2008}, here upper limit of summation are chosen to as $3$. 
The results presented in these tables show that by the use of STSOs the basis sets can be extended with arbitrary sign of $\kappa$ and the optimization procedure can be applied to obtain minimum values of desired energy states without encountering variational instability. 
For low nuclear charges during the optimization of the screening constants a particular energy states proceed towards the gap. 
It can be seen from the figure plotted for these energy states depending screening constants these exceptions can be neglected as the spurious roots encountered far from minimum values. 
\\
The method is also tested for higher uppper limit of summation in table \ref{tab:table4}. Here, the given results in second row of each nuclear charges are the calculations performed with nonrelativistic limit. It is observed that for low values of nuclear charges it may require to take into account one more condition; different from nonrelativistic case in relativistic calculations the eigenvalues obtained by solution of Dirac equation is not in general an upper bounds while Kato's upper and lower bounds given as \cite{Chen1994},
\begin{align} \label{eq:KATOUPLOW}
E_{low} \leqslant E_{i} \leqslant E_{up}
\end{align}
here $E_{i}$ are eigenvalues of spectrum obtained by the solution of Dirac equation which suppose in interval $(a,b)$;
\begin{align} \label{eq:KATO}
a<\langle H_{D} \rangle<b,
\end{align}
\begin{align} \label{eq:KATOLOW}
E_{low}=
\langle H_{D} \rangle-
\frac{s^{2}}{b-\langle H_{D} \rangle}>a,
\end{align}
\begin{align} \label{eq:KATOUP}
E_{up}=
\langle H_{D} \rangle+
\frac{s^{2}}{\langle H_{D} \rangle-a}<b,
\end{align}
where,
\begin{align} \label{eq:ROOT-MEAN-SQUARE}
s
=\sqrt{\left(\langle H_{D}^{2} \rangle-\langle H_{D} \rangle^{2} \right)}.
\end{align}
Some results in table \ref{tab:table4} should consider depending on this condition. Note that calculation of root-mean-square deviation s is not easy in the case of two-center problem. It may possible to find practical method on this issue. For now, it is outside the scobe of this paper. 

The results of calculations performed by literature specially were carried out for $H_{2}^{+}$, $HeH^{2+}$, $Sn_{2}^{99+}$, $Th_{2}^{179+}$ \cite{Mark1980, Quiney2004, LaJohn1992, Laaksonen1984, Yang1991, Yu1994, Alexander1999, Kullie2001, Ishikawa2008, Artemyev2010, Mark1987, Howells1990, Franke1992, Parpia1995, Tsogbayar2006} by setting the internuclear distances $R/Z$ and $R=2a.u.$ for $Sn_{2}^{99+}$, $Th_{2}^{179+}$ molecules.
\\
Note that in this study all results are given in atomic units (a.u.).

\section*{Acknowledgement}
A.B. acknowledges funding for a postdoctoral research fellowship from innov@pole: the Auvergne Region and FEDER. 

The tables given in this study are the first version of calculations and it is believed that they will be helpful on behalf of the reader's views. In the next version of the paper they will be revised. Presentation will be just for highest upper limit of summation considering upper and lower bounds in relativistic theory.

\begin{table*}
\caption{\label{tab:table1} Electronic energies for the ground and excited states of some one-electron diatomic molecules using minimal basis sets with internuclear distances $R=2,5,10$ for each symmetry.}
\begin{ruledtabular}
\begin{tabular}{cccccccccc}
 $Z_{a};Z_{b}$ & \multicolumn{3}{c}{$1s\sigma_{1/2}$} & \multicolumn{3}{c}{$1p\pi_{3/2}$} & \multicolumn{3}{c}{$2d\delta_{5/2}$} 
\\ 
\hline
$1;1$
& \begin{tabular}[c]{@{}l@{}}1.086514728\\ 1.086505992\end{tabular} 
& \begin{tabular}[c]{@{}l@{}}0.719210770\\ 0.719205456\end{tabular} 
& \begin{tabular}[c]{@{}l@{}}0.600304482\\ 0.600297890\end{tabular} 
& \begin{tabular}[c]{@{}l@{}}0.425906666\\ 0.425907162\end{tabular}
& \begin{tabular}[c]{@{}l@{}}0.312887102\\ 0.312887451\end{tabular}
& \begin{tabular}[c]{@{}l@{}}0.226879709\\ 0.226879635\end{tabular} 
& \begin{tabular}[c]{@{}l@{}}0.212459788\\ 0.202958953\end{tabular} 
& \begin{tabular}[c]{@{}l@{}}0.182364551\\ 0.153864857\end{tabular} 
& \begin{tabular}[c]{@{}l@{}}0.142114476\\ 0.120799851\end{tabular}  
\\
$1;2$
& \begin{tabular}[c]{@{}l@{}}2.504460112\\ 2.504352324\end{tabular} 
& \begin{tabular}[c]{@{}l@{}}2.200106708\\ 2.200000193\end{tabular} 
& \begin{tabular}[c]{@{}l@{}}2.100106514\\ 2.100000000\end{tabular} 
& \begin{tabular}[c]{@{}l@{}}0.890914639\\ 0.890911570\end{tabular} 
& \begin{tabular}[c]{@{}l@{}}0.689319577\\ 0.689312571\end{tabular} 
& \begin{tabular}[c]{@{}l@{}}0.598511770\\ 0.598505033\end{tabular} 
& \begin{tabular}[c]{@{}l@{}}0.461732618\\ 0.422805097\end{tabular} 
& \begin{tabular}[c]{@{}l@{}}0.380247382\\ 0.324979675\end{tabular} 
& \begin{tabular}[c]{@{}l@{}}0.314347999\\ 0.304904143\end{tabular}                              
\\
$25;25$
& \begin{tabular}[c]{@{}l@{}}327.6443552\\ 325.0000000\end{tabular} 
& \begin{tabular}[c]{@{}l@{}}320.1443552\\ 317.5000000\end{tabular} 
& \begin{tabular}[c]{@{}l@{}}317.6443552\\ 315.0000000\end{tabular} 
& \begin{tabular}[c]{@{}l@{}}90.75845109\\ 90.59501178\end{tabular} 
& \begin{tabular}[c]{@{}l@{}}83.28628590\\ 83.12308004\end{tabular} 
& \begin{tabular}[c]{@{}l@{}}80.78795188\\ 80.62476000\end{tabular} 
& \begin{tabular}[c]{@{}l@{}}47.07944520\\ 47.00480016\end{tabular} 
& \begin{tabular}[c]{@{}l@{}}39.74293357\\ 39.71029603\end{tabular} 
& \begin{tabular}[c]{@{}l@{}}37.25294866\\ 37.22076945\end{tabular} 
\\
$25;30$
& \begin{tabular}[c]{@{}l@{}}468.0249072\\ 462.5000000\end{tabular} 
& \begin{tabular}[c]{@{}l@{}}460.5249071\\ 455.0000000\end{tabular} 
& \begin{tabular}[c]{@{}l@{}}458.0249071\\ 452.5000000\end{tabular} 
& \begin{tabular}[c]{@{}l@{}}125.3184354\\ 124.9791705\end{tabular} 
& \begin{tabular}[c]{@{}l@{}}117.8376980\\ 117.4986667\end{tabular} 
& \begin{tabular}[c]{@{}l@{}}115.3388506\\ 114.9998333\end{tabular} 
& \begin{tabular}[c]{@{}l@{}}62.44386351\\ 62.35619384\end{tabular} 
& \begin{tabular}[c]{@{}l@{}}55.05879776\\ 54.99180341\end{tabular} 
& \begin{tabular}[c]{@{}l@{}}52.56574774\\ 52.49899383\end{tabular} 
\\
$50;50$
& \begin{tabular}[c]{@{}l@{}}1319.626156\\ 1275.000000\end{tabular} 
& \begin{tabular}[c]{@{}l@{}}1304.626156\\ 1260.000000\end{tabular} 
& \begin{tabular}[c]{@{}l@{}}1299.626156\\ 1255.000000\end{tabular} 
& \begin{tabular}[c]{@{}l@{}}340.1298551\\ 337.4850007\end{tabular} 
& \begin{tabular}[c]{@{}l@{}}325.1434271\\ 322.4990400\end{tabular} 
& \begin{tabular}[c]{@{}l@{}}320.1442392\\ 317.4998799\end{tabular} 
& \begin{tabular}[c]{@{}l@{}}164.3180180\\ 163.7939488\end{tabular} 
& \begin{tabular}[c]{@{}l@{}}149.4006642\\ 148.8830778\end{tabular} 
& \begin{tabular}[c]{@{}l@{}}144.4056265\\ 143.8881673\end{tabular} 
\\
$50;60$
& \begin{tabular}[c]{@{}l@{}}1920.682356\\ 1825.000000\end{tabular} 
& \begin{tabular}[c]{@{}l@{}}1905.682356\\ 1810.000000\end{tabular} 
& \begin{tabular}[c]{@{}l@{}}1900.682356\\ 1805.000000\end{tabular} 
& \begin{tabular}[c]{@{}l@{}}480.5149899\\ 474.9895836\end{tabular} 
& \begin{tabular}[c]{@{}l@{}}465.5242724\\ 459.9993333\end{tabular} 
& \begin{tabular}[c]{@{}l@{}}460.5248278\\ 454.9999167\end{tabular} 
& \begin{tabular}[c]{@{}l@{}}226.0155137\\ 222.9362638\end{tabular} 
& \begin{tabular}[c]{@{}l@{}}211.0726101\\ 209.9959753\end{tabular} 
& \begin{tabular}[c]{@{}l@{}}206.0760341\\ 204.9994992\end{tabular} 
\\
$100;100$
& \begin{tabular}[c]{@{}l@{}}5989.195384\\ 5050.000000\end{tabular} 
& \begin{tabular}[c]{@{}l@{}}5959.195384\\ 5020.000000\end{tabular} 
& \begin{tabular}[c]{@{}l@{}}5949.195384\\ 5010.000000\end{tabular} 
& \begin{tabular}[c]{@{}l@{}}1344.619654\\ 1299.992500\end{tabular} 
& \begin{tabular}[c]{@{}l@{}}1314.625740\\ 1269.999520\end{tabular} 
& \begin{tabular}[c]{@{}l@{}}1304.626104\\ 1259.999939\end{tabular} 
& \begin{tabular}[c]{@{}l@{}}613.9835542\\ 605.5099329\end{tabular} 
& \begin{tabular}[c]{@{}l@{}}584.0231455\\ 575.5526691\end{tabular} 
& \begin{tabular}[c]{@{}l@{}}574.0255160\\ 565.5551954\end{tabular} 
\\
$100;1$
& \begin{tabular}[c]{@{}l@{}}5939.695384\\ 5000.499999\end{tabular} 
& \begin{tabular}[c]{@{}l@{}}5939.395384\\ 5000.200000\end{tabular} 
& \begin{tabular}[c]{@{}l@{}}5939.295384\\ 5000.100000\end{tabular} 
& \begin{tabular}[c]{@{}l@{}}1295.126091\\ 1250.499925\end{tabular} 
& \begin{tabular}[c]{@{}l@{}}1294.826152\\ 1250.199995\end{tabular} 
& \begin{tabular}[c]{@{}l@{}}1294.726155\\ 1250.099999\end{tabular} 
& \begin{tabular}[c]{@{}l@{}}564.5254317\\ 556.0550992\end{tabular} 
& \begin{tabular}[c]{@{}l@{}}564.2258276\\ 555.7555266\end{tabular} 
& \begin{tabular}[c]{@{}l@{}}564.1258513\\ 555.6555520\end{tabular} 
\\
\end{tabular}
\end{ruledtabular}
\end{table*}

\begin{table*}
\caption{\label{tab:table2}Electronic energies for the ground and excited states of some one-electron diatomic molecules using extended basis sets approximation where, $N_{q}=1$ with internuclear distances $R=2,5,10$ for each symmetry.}
\begin{ruledtabular}
\begin{tabular}{cccccccccc}
 $Z_{a};Z_{b}$ & \multicolumn{3}{c}{$1s\sigma_{1/2}$} & \multicolumn{3}{c}{$1p\pi_{3/2}$} & \multicolumn{3}{c}{$2d\delta_{5/2}$} 
\\ 
\hline
$1;1$
& \begin{tabular}[c]{@{}l@{}}1.091130119\\ 1.091120822\end{tabular} 
& \begin{tabular}[c]{@{}l@{}}0.720765169\\ 0.720760135\end{tabular} 
& \begin{tabular}[c]{@{}l@{}}0.600393466\\ 0.600078770\end{tabular}
& \begin{tabular}[c]{@{}l@{}}0.426273945\\ 0.426274153\end{tabular}
& \begin{tabular}[c]{@{}l@{}}0.314446554\\ 0.314450113\end{tabular}
& \begin{tabular}[c]{@{}l@{}}0.227888953\\ 0.227888279\end{tabular} 
& \begin{tabular}[c]{@{}l@{}}0.212473037\\ 0.212471077\end{tabular}
& \begin{tabular}[c]{@{}l@{}}0.182708840\\ 0.182707620\end{tabular} 
& \begin{tabular}[c]{@{}l@{}}0.142861983\\ 0.142865469\end{tabular}  
\\
$1;2$
& \begin{tabular}[c]{@{}l@{}}2.506703860\\ 2.506554826\end{tabular} 
& \begin{tabular}[c]{@{}l@{}}2.200181554\\ 2.200085087\end{tabular} 
& \begin{tabular}[c]{@{}l@{}}2.100111174\\ 2.100020055\end{tabular} 
& \begin{tabular}[c]{@{}l@{}}0.892330988\\ 0.892327358\end{tabular} 
& \begin{tabular}[c]{@{}l@{}}0.690154901\\ 0.690103120\end{tabular} 
& \begin{tabular}[c]{@{}l@{}}0.598595445\\ 0.598588726\end{tabular} 
& \begin{tabular}[c]{@{}l@{}}0.461861897\\ 0.461811508\end{tabular} 
& \begin{tabular}[c]{@{}l@{}}0.381038054\\ 0.381034503\end{tabular} 
& \begin{tabular}[c]{@{}l@{}}0.314643291\\ 0.314664141\end{tabular}                              
\\
$25;25$
& \begin{tabular}[c]{@{}l@{}}327.6444253\\ 325.0000825\end{tabular} 
& \begin{tabular}[c]{@{}l@{}}320.1443570\\ 317.5000379\end{tabular} 
& \begin{tabular}[c]{@{}l@{}}317.6443553\\ 315.0002009\end{tabular} 
& \begin{tabular}[c]{@{}l@{}}90.75994619\\ 90.59668066\end{tabular} 
& \begin{tabular}[c]{@{}l@{}}83.28632497\\ 83.12312128\end{tabular} 
& \begin{tabular}[c]{@{}l@{}}80.78795429\\ 80.62476195\end{tabular} 
& \begin{tabular}[c]{@{}l@{}}47.08799551\\ 47.05497574\end{tabular} 
& \begin{tabular}[c]{@{}l@{}}39.74317806\\ 39.71097697\end{tabular} 
& \begin{tabular}[c]{@{}l@{}}37.25296417\\ 37.22079863\end{tabular} 
\\
$25;30$
& \begin{tabular}[c]{@{}l@{}}468.0249399\\ 462.5000450\end{tabular} 
& \begin{tabular}[c]{@{}l@{}}460.5249080\\ 455.0002005\end{tabular} 
& \begin{tabular}[c]{@{}l@{}}458.0249072\\ 452.5000001\end{tabular} 
& \begin{tabular}[c]{@{}l@{}}125.3191570\\ 124.9799087\end{tabular} 
& \begin{tabular}[c]{@{}l@{}}117.8377167\\ 117.4986809\end{tabular} 
& \begin{tabular}[c]{@{}l@{}}115.3388518\\ 114.9997025\end{tabular} 
& \begin{tabular}[c]{@{}l@{}}62.44812897\\ 62.38072695\end{tabular} 
& \begin{tabular}[c]{@{}l@{}}55.05891605\\ 54.99213261\end{tabular} 
& \begin{tabular}[c]{@{}l@{}}52.56575522\\ 52.49900475\end{tabular} 
\\
$50;50$
& \begin{tabular}[c]{@{}l@{}}1319.626169\\ 1275.000019\end{tabular} 
& \begin{tabular}[c]{@{}l@{}}1304.626156\\ 1260.000036\end{tabular} 
& \begin{tabular}[c]{@{}l@{}}1299.626156\\ 1255.000001\end{tabular} 
& \begin{tabular}[c]{@{}l@{}}340.1302179\\ 337.4854005\end{tabular} 
& \begin{tabular}[c]{@{}l@{}}325.1434365\\ 322.4991611\end{tabular} 
& \begin{tabular}[c]{@{}l@{}}320.1442396\\ 317.4998801\end{tabular} 
& \begin{tabular}[c]{@{}l@{}}164.3203270\\ 163.8016252\end{tabular} 
& \begin{tabular}[c]{@{}l@{}}149.4007250\\ 148.8831807\end{tabular} 
& \begin{tabular}[c]{@{}l@{}}144.4056303\\ 143.8881692\end{tabular} 
\\
$50;60$
& \begin{tabular}[c]{@{}l@{}}1920.682362\\ 1825.000325\end{tabular} 
& \begin{tabular}[c]{@{}l@{}}1905.682357\\ 1810.000002\end{tabular} 
& \begin{tabular}[c]{@{}l@{}}1900.682356\\ 1805.000033\end{tabular} 
& \begin{tabular}[c]{@{}l@{}}480.5151602\\ 474.9897706\end{tabular} 
& \begin{tabular}[c]{@{}l@{}}465.5242768\\ 459.9992962\end{tabular} 
& \begin{tabular}[c]{@{}l@{}}460.5248280\\ 454.9999089\end{tabular} 
& \begin{tabular}[c]{@{}l@{}}226.0166250\\ 224.9388235\end{tabular} 
& \begin{tabular}[c]{@{}l@{}}211.0726391\\ 209.9960260\end{tabular} 
& \begin{tabular}[c]{@{}l@{}}206.0760351\\ 204.9994913\end{tabular} 
\\
$100;100$
& \begin{tabular}[c]{@{}l@{}}5989.195386\\ 5050.000026\end{tabular} 
& \begin{tabular}[c]{@{}l@{}}5959.195384\\ 5020.000013\end{tabular}
& \begin{tabular}[c]{@{}l@{}}5949.195384\\ 5010.000013\end{tabular} 
& \begin{tabular}[c]{@{}l@{}}1344.619729\\ 1299.992597\end{tabular} 
& \begin{tabular}[c]{@{}l@{}}1314.625742\\ 1269.999693\end{tabular} 
& \begin{tabular}[c]{@{}l@{}}1304.626104\\ 1259.999941\end{tabular} 
& \begin{tabular}[c]{@{}l@{}}613.9840961\\ 605.5112314\end{tabular} 
& \begin{tabular}[c]{@{}l@{}}584.0231594\\ 575.5528592\end{tabular} 
& \begin{tabular}[c]{@{}l@{}}574.0255152\\ 565.5553282\end{tabular} 
\\
$100;1$
& \begin{tabular}[c]{@{}l@{}}5939.695384\\ 5000.500012\end{tabular}
& \begin{tabular}[c]{@{}l@{}}5939.395396\\ 5000.200012\end{tabular} 
& \begin{tabular}[c]{@{}l@{}}5939.295384\\ 5000.100004\end{tabular} 
& \begin{tabular}[c]{@{}l@{}}1295.126091\\ 1250.499906\end{tabular} 
& \begin{tabular}[c]{@{}l@{}}1294.826152\\ 1250.199996\end{tabular} 
& \begin{tabular}[c]{@{}l@{}}1294.726155\\ 1250.099820\end{tabular} 
& \begin{tabular}[c]{@{}l@{}}564.5254317\\ 556.0551075\end{tabular} 
& \begin{tabular}[c]{@{}l@{}}564.2258276\\ 555.7555269\end{tabular} 
& \begin{tabular}[c]{@{}l@{}}564.1258513\\ 555.6555521\end{tabular} 
\\
\end{tabular}
\end{ruledtabular}
\end{table*}

\begin{table*}
\caption{\label{tab:table3}Electronic energies for the ground and excited states of some one-electron diatomic molecules using extended basis sets approximation where, $N_{q}=2$ with internuclear distances $R=2,5,10$ for each symmetry.}
\begin{ruledtabular}
\begin{tabular}{cccccccccc}
 $Z_{a};Z_{b}$ & \multicolumn{3}{c}{$1s\sigma_{1/2}$} & \multicolumn{3}{c}{$1p\pi_{3/2}$} & \multicolumn{3}{c}{$2d\delta_{5/2}$}  
\\ 
\hline
$1;1$
& \begin{tabular}[c]{@{}l@{}}1.101601204           \\ 1.102189453\end{tabular} 
& \begin{tabular}[c]{@{}l@{}}0.725110669           \\ 0.723828536\end{tabular} 
& \begin{tabular}[c]{@{}l@{}}0.600603136           \\ 0.600559514\end{tabular} 
& \begin{tabular}[c]{@{}l@{}}0.429282890           \\ 0.428734453\end{tabular} 
& \begin{tabular}[c]{@{}l@{}}\multicolumn{1}{c}-   \\ 0.320682934\end{tabular} 
& \begin{tabular}[c]{@{}l@{}}\multicolumn{1}{c}-   \\ 0.231879320\end{tabular} 
& \begin{tabular}[c]{@{}l@{}}0.227487111           \\ 0.204241193\end{tabular} 
& \begin{tabular}[c]{@{}l@{}}0.187385304           \\ 0.159462805\end{tabular} 
& \begin{tabular}[c]{@{}l@{}}\multicolumn{1}{c}-   \\ 0.124600244\end{tabular}  
\\
$1;2$
& \begin{tabular}[c]{@{}l@{}}2.512050427           \\ 2.511648703\end{tabular} 
& \begin{tabular}[c]{@{}l@{}}2.200341532           \\ 2.200225510\end{tabular} 
& \begin{tabular}[c]{@{}l@{}}2.100120217           \\ 2.100013995\end{tabular} 
& \begin{tabular}[c]{@{}l@{}}0.900534611           \\ 0.899513765\end{tabular} 
& \begin{tabular}[c]{@{}l@{}}\multicolumn{1}{c}-   \\ 0.693759040\end{tabular} 
& \begin{tabular}[c]{@{}l@{}}0.599035320           \\ 0.598930437\end{tabular} 
& \begin{tabular}[c]{@{}l@{}}0.464719295           \\ 0.429056911\end{tabular} 
& \begin{tabular}[c]{@{}l@{}}\multicolumn{1}{c}-   \\ 0.342212791\end{tabular} 
& \begin{tabular}[c]{@{}l@{}}\multicolumn{1}{c}-   \\ 0.305794455\end{tabular}                              
\\
$25;25$
& \begin{tabular}[c]{@{}l@{}}327.6445471\\ 325.0002238\end{tabular} 
& \begin{tabular}[c]{@{}l@{}}320.1443592\\ 317.5000057\end{tabular} 
& \begin{tabular}[c]{@{}l@{}}317.6443555\\ 315.0000004\end{tabular} 
& \begin{tabular}[c]{@{}l@{}}90.76701622\\ 90.60262775\end{tabular} 
& \begin{tabular}[c]{@{}l@{}}83.28648284\\ 83.12327854\end{tabular} 
& \begin{tabular}[c]{@{}l@{}}80.78796374\\ 80.62477244\end{tabular} 
& \begin{tabular}[c]{@{}l@{}}47.18828465\\ 47.05909953\end{tabular} 
& \begin{tabular}[c]{@{}l@{}}39.74489281\\ 39.71201767\end{tabular} 
& \begin{tabular}[c]{@{}l@{}}37.25306098\\ 37.22087892\end{tabular} 
\\
$25;30$
& \begin{tabular}[c]{@{}l@{}}468.0249870\\ 462.5001079\end{tabular} 
& \begin{tabular}[c]{@{}l@{}}460.5249092\\ 455.0000028\end{tabular} 
& \begin{tabular}[c]{@{}l@{}}458.0249073\\ 452.5000002\end{tabular} 
& \begin{tabular}[c]{@{}l@{}}125.3222987\\ 124.9828671\end{tabular} 
& \begin{tabular}[c]{@{}l@{}}117.8377902\\ 117.4987625\end{tabular} 
& \begin{tabular}[c]{@{}l@{}}115.3388562\\ 114.9998393\end{tabular} 
& \begin{tabular}[c]{@{}l@{}}62.48547808\\ 62.38497436\end{tabular} 
& \begin{tabular}[c]{@{}l@{}}55.05968700\\ 54.99263991\end{tabular} 
& \begin{tabular}[c]{@{}l@{}}52.56580039\\ 52.49904670\end{tabular} 
\\
$50;50$
& \begin{tabular}[c]{@{}l@{}}1319.626218\\ 1275.000056\end{tabular} 
& \begin{tabular}[c]{@{}l@{}}1304.626161\\ 1260.000001\end{tabular} 
& \begin{tabular}[c]{@{}l@{}}1299.626157\\ 1255.000000\end{tabular} 
& \begin{tabular}[c]{@{}l@{}}340.1317069\\ 337.4869354\end{tabular} 
& \begin{tabular}[c]{@{}l@{}}325.1434719\\ 322.4990898\end{tabular} 
& \begin{tabular}[c]{@{}l@{}}320.1442424\\ 317.4998831\end{tabular} 
& \begin{tabular}[c]{@{}l@{}}164.3370163\\ 163.8105069\end{tabular} 
& \begin{tabular}[c]{@{}l@{}}149.4011045\\ 148.8835157\end{tabular} 
& \begin{tabular}[c]{@{}l@{}}144.4056531\\ 143.8881948\end{tabular} 
\\
$50;60$
& \begin{tabular}[c]{@{}l@{}}1920.682417\\ 1825.000027\end{tabular} 
& \begin{tabular}[c]{@{}l@{}}1905.682364\\ 1810.000001\end{tabular} 
& \begin{tabular}[c]{@{}l@{}}1900.682358\\ 1805.000000\end{tabular} 
& \begin{tabular}[c]{@{}l@{}}480.5158274\\ 474.9905181\end{tabular} 
& \begin{tabular}[c]{@{}l@{}}465.5242956\\ 459.9993573\end{tabular} 
& \begin{tabular}[c]{@{}l@{}}460.5248301\\ 454.9999182\end{tabular} 
& \begin{tabular}[c]{@{}l@{}}226.0241758\\ 224.9431969\end{tabular} 
& \begin{tabular}[c]{@{}l@{}}211.0728152\\ 209.9961868\end{tabular} 
& \begin{tabular}[c]{@{}l@{}}206.0760468\\ 204.9995125\end{tabular} 
\\
$100;100$
& \begin{tabular}[c]{@{}l@{}}5989.196040\\ 5050.000014\end{tabular} 
& \begin{tabular}[c]{@{}l@{}}5959.195489\\ 5020.000001\end{tabular} 
& \begin{tabular}[c]{@{}l@{}}5949.195410\\ 5010.000000\end{tabular} 
& \begin{tabular}[c]{@{}l@{}}1344.620314\\ 1299.992986\end{tabular} 
& \begin{tabular}[c]{@{}l@{}}1314.625803\\ 1269.999532\end{tabular} 
& \begin{tabular}[c]{@{}l@{}}1304.626118\\ 1259.999941\end{tabular} 
& \begin{tabular}[c]{@{}l@{}}613.9876542\\ 605.5141963\end{tabular} 
& \begin{tabular}[c]{@{}l@{}}584.0232703\\ 575.5527790\end{tabular} 
& \begin{tabular}[c]{@{}l@{}}574.0255302\\ 565.5552022\end{tabular} 
\\
$1;100$
& \begin{tabular}[c]{@{}l@{}}5939.695384\\ 5000.500000\end{tabular} 
& \begin{tabular}[c]{@{}l@{}}5939.395384\\ 5000.199999\end{tabular} 
& \begin{tabular}[c]{@{}l@{}}5939.295384\\ 5000.099999\end{tabular} 
& \begin{tabular}[c]{@{}l@{}}1295.126091\\ 1250.499925\end{tabular} 
& \begin{tabular}[c]{@{}l@{}}1294.826152\\ 1250.199995\end{tabular} 
& \begin{tabular}[c]{@{}l@{}}1294.726155\\ 1250.099999\end{tabular} 
& \begin{tabular}[c]{@{}l@{}}564.5254320\\ 556.0550997\end{tabular} 
& \begin{tabular}[c]{@{}l@{}}564.2258276\\ 555.7555267\end{tabular} 
& \begin{tabular}[c]{@{}l@{}}564.1258513\\ 555.6555520\end{tabular} 
\\
\end{tabular}
\end{ruledtabular}
\begin{flushleft}
R=2: $Z_{a};Z_{b}$=1;1, $1s\sigma_{1/2}$: 1.10131\cite{Mark1980},1.1026415801\cite{Parpia1995},1.1026411255\cite{Ishikawa2008}\hspace{10pt} $Z_{a};Z_{b}$=1;2, $1s\sigma_{1/2}$: 2.512296099\cite{Laaksonen1984} 
\\
R=2: $Z_{a};Z_{b}$=1;1, $1s\sigma_{1/2}$: 1.10232$\left(c\rightarrow \infty\right)$\cite{Mark1980},1.1026342137$\left(c\rightarrow 10^{3}\right)$\cite{Parpia1995}\hspace{6pt} $Z_{a};Z_{b}$=1;2, $1s\sigma_{1/2}$: 2.512193020$\left(c\rightarrow 10^{7}\right)$\cite{Laaksonen1984} \\
R=2: $Z_{a};Z_{b}$=1;1, $1s\sigma$: 1.1026342145\cite{Ishikawa2008} \hspace{137pt} $Z_{a};Z_{b}$=1;2, $1s\sigma$: 2.512192938\cite{Bagci2008}
\\
R=5: $Z_{a};Z_{b}$=1;1, $1s\sigma$: 0.7244202952\cite{Campos2006} \hspace{137pt} $Z_{a};Z_{b}$=1;2, $1s\sigma$: 2.200236963\cite{Bagci2008} \\
R=10: $Z_{a};Z_{b}$=1;1, $1s\sigma$: 0.6005787289\cite{Campos2006} \hspace{132pt} $Z_{a};Z_{b}$=1;2, $1s\sigma$: 2.100014205\cite{Bagci2008}
\\
R=2: $Z_{a};Z_{b}$=1;1, $2p\pi$: 0.428770950\cite{Franke1992} \hspace{141pt} $Z_{a};Z_{b}$=1;2, $2p\pi$: 0.899646586\cite{Franke1992}, 0.899646663\cite{Bagci2008}
\\
R=5: - \hspace{272pt} $Z_{a};Z_{b}$=1;2, $2p\pi$: 0.694221941\cite{Bagci2008}
\\
R=10: - \hspace{267pt} $Z_{a};Z_{b}$=1;2, $2p\pi$: 0.598963473\cite{Bagci2008}
\\
R=2: - \hspace{272pt} $Z_{a};Z_{b}$=1;2, $3d\delta$: 0.463296596\cite{Bagci2008}
\\
R=5: - \hspace{272pt} $Z_{a};Z_{b}$=1;2, $3d\delta$: 0.386333590\cite{Bagci2008}
\\
R=10: - \hspace{267pt} $Z_{a};Z_{b}$=1;2, $3d\delta$: 0.316846057\cite{Bagci2008}
\end{flushleft}
\end{table*}

\begin{table}
\caption{\label{tab:table4}Electronic energies for the ground and excited states of some one-electron diatomic molecules using extended basis sets approximation where, $N_{q}=3,4$ and internuclear distance $R=2$.}
\begin{ruledtabular}
\begin{tabular}{cccccc}
 $Z_{a};Z_{b}$ & \multicolumn{2}{c}{$\vert \kappa \vert =3$} & $\vert \kappa \vert =4$
\\
& $1s\sigma_{1/2}$ & $1p\pi_{3/2}$ & $1s\sigma_{1/2}$
\\ 
\hline
$1;1$
& \begin{tabular}[c]{@{}l@{}}1.102372222\\ 1.102357336\end{tabular} 
& \begin{tabular}[c]{@{}l@{}}0.453060085\\ 0.453045512\end{tabular} 
& \begin{tabular}[c]{@{}l@{}}1.104011504\\ 1.103989240\end{tabular} 
\\
$1;2$
& \begin{tabular}[c]{@{}l@{}}\multicolumn{1}{c}-\\ \multicolumn{1}{c}-\end{tabular} 
& \begin{tabular}[c]{@{}l@{}}0.915679457\\ 0.915222994\end{tabular} 
& \begin{tabular}[c]{@{}l@{}}2.512411017\\ 2.512119450\end{tabular} 
\\
$25;25$
& \begin{tabular}[c]{@{}l@{}}327.6445472\\ 324.9094337\end{tabular} 
& \begin{tabular}[c]{@{}l@{}}90.76709063\\ 90.60246130\end{tabular}  
& \begin{tabular}[c]{@{}l@{}}327.6445472\\ 324.9094076\end{tabular} 
\\
$25;30$
& \begin{tabular}[c]{@{}l@{}}468.0249870\\ 462.2217421\end{tabular} 
& \begin{tabular}[c]{@{}l@{}}125.3223169\\ 124.9791134\end{tabular} 
& \begin{tabular}[c]{@{}l@{}}468.0249870\\ 462.2217782\end{tabular} 
\\
$50;50$
& \begin{tabular}[c]{@{}l@{}}\multicolumn{1}{c}-\\ 2755.892565\end{tabular} 
& \begin{tabular}[c]{@{}l@{}}\multicolumn{1}{c}-\\ 1065.686764\footnotemark[1] \end{tabular}
& \begin{tabular}[c]{@{}l@{}}\multicolumn{1}{c}-\\ 2759.976757\end{tabular}
\\
$90;90$
& \begin{tabular}[c]{@{}l@{}}\multicolumn{1}{c}-\\ 8929.091938\end{tabular} 
& \begin{tabular}[c]{@{}l@{}}\multicolumn{1}{c}-\\ 3449.848015\footnotemark[1] \end{tabular}
& \begin{tabular}[c]{@{}l@{}}\multicolumn{1}{c}-\\ 8942.324615\end{tabular}

\end{tabular}
\end{ruledtabular}
\begin{flushleft}
$Z_{a};Z_{b}$=50;50, $1s\sigma_{1/2}$: 2807.25\cite{Mark1980} \hspace{20pt} $1s\sigma$: 2756.59\cite{Mark1980}
\\
$Z_{a};Z_{b}$=90;90, $1s\sigma_{1/2}$: 9496.04\cite{Mark1980} \hspace{20pt} $1s\sigma$: 8931.35\cite{Mark1980}
\\
$Z_{a};Z_{b}$=90;90, $1s\sigma_{1/2} \left(c\times 10^{3} \right)$: 9504.756696\cite{Parpia1995}
\\
$Z_{a};Z_{b}$=90;90, $1s\sigma$: 8931.337137\cite{Parpia1995}
\\
$Z_{a};Z_{b}$=50;50, $2p\pi$: 1064.767903(Minimal basis sets)
\\
$Z_{a};Z_{b}$=90;90, $2p\pi$: 3449.848014(Minimal basis sets)
\end{flushleft}
\footnotetext[1]{Minimal basis sets}
\end{table}

\begin{figure*}
\includegraphics[width=1.00\textwidth,height=0.50\textheight]{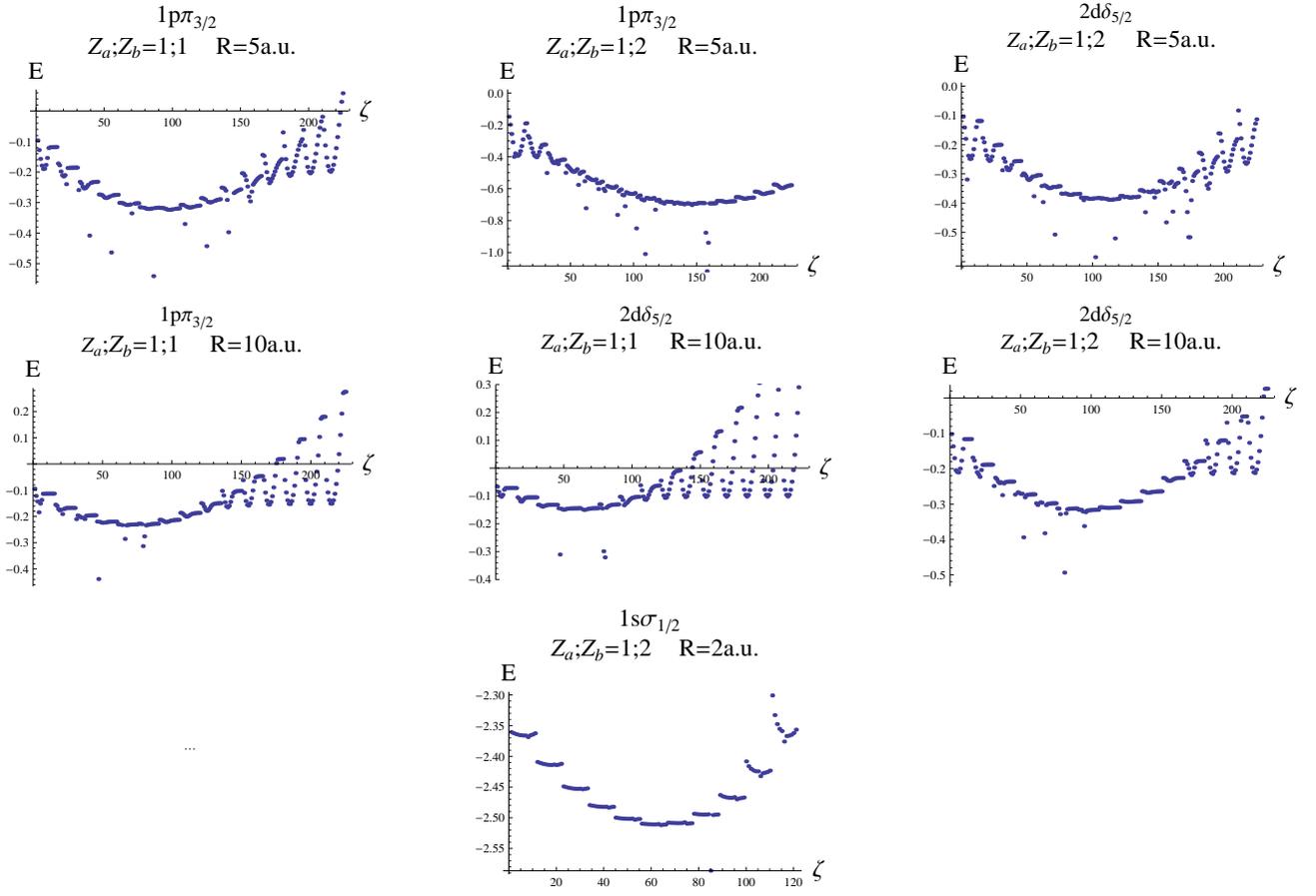}% Here is how to import EPS art
\caption{\label{fig:ESTATES} Dependence of electronic energy states to screening constants.}
\end{figure*}

\end{document}